\documentclass{ieee}

\usepackage{times}
\usepackage{epsf}

\newcommand{\be}{\begin{equation}}
\newcommand{\ee}{\end{equation}}
\newcommand{\ba}{\begin{eqnarray}}
\newcommand{\ea}{\end{eqnarray}}
\newcommand{\Tr}{\mbox{Tr}}
\newcommand{\1}{1\!\!\!\bot}
\def\reff#1{(\ref{#1})}

\def\bun{e}
\def\bx{x}
\def\by{y}
\def\bum{\vec{\mbox{$A$}}}
\def\bk{k}
\def\bfv{\vec{\mbox{$f$}}}

\def\bsig{\vec{\mbox{$\sigma$}}}
\def\bvg{\vec{\mbox{$g$}}}
\def\bvw{\vec{\mbox{${\mathrm w}$}}}
\def\bvu{\vec{\mbox{$u$}}}
\def\bvv{\vec{\mbox{$v$}}}

\begin{document}

\title{Parallel implementation of a lattice-gauge-theory
       code: \\ studying quark confinement on PC clusters} 

\author{Attilio Cucchieri, Tereza Mendes, Gonzalo Travieso\\
        Instituto de F\'\i sica de S\~ao Carlos,
        Universidade de S\~ao Paulo \\
        C.P.\ 369, 13560-970 S\~ao Carlos, SP, Brazil \\
        attilio@if.sc.usp.br, mendes@if.sc.usp.br,
        gonzalo@if.sc.usp.br\\
        \and
        Andre R.\ Taurines\\
        Instituto de F\'\i sica,
        Universidade Federal do Rio Grande do Sul \\
        Av.\ Bento Gon\c calves, 9500, Campus do Vale,
        91501-970 Porto Alegre, RS, Brazil \\
        taurines@if.ufrgs.br\\
}

\maketitle
\thispagestyle{empty}

\begin{abstract}
We consider the implementation of a parallel Monte Carlo
code for high-performance simulations on PC clusters with MPI.
We carry out tests of speedup and efficiency.
The code is used for numerical simulations of
pure $SU(2)$ lattice gauge theory at
very large lattice volumes, in order to study the
infrared behavior of gluon and ghost propagators. This
problem is directly related to the confinement of quarks
and gluons in the physics of strong interactions.
\end{abstract}


\section{Introduction}

The strong force is one of the four fundamental interactions
of nature (along with gravity, electromagnetism and the
weak force). It is the force that holds together protons
and neutrons in the
nucleus. The strong interaction is described by Quantum Chromodynamics
(QCD), a quantum field theory with local $SU(3)$ gauge invariance
\cite{moriyasu}.
QCD states that a baryon (e.g.\ a proton or a neutron) is not an
elementary particle but is instead made up of building blocks
called {\bf quarks}, interacting through the exchange of massless
particles called {\bf gluons} (equivalent to the photons in the
electromagnetic interaction).
A unique feature of the strong force is that the particles that feel it
directly --- quarks and gluons --- are completely hidden from us,
i.e.\ they are never observed as free particles.
This property is known as {\bf confinement} and makes
QCD much harder to handle theoretically than the theories
describing the weak and electromagnetic forces.
Indeed, it is not possible to study the confinement problem
analytically and physicists must therefore rely on numerical
simulations performed on supercomputers. These studies
are done using the {\bf lattice} formulation of QCD \cite{rothe},
which is based on field quantization through Feynman
integrals and discretization
of space-time on a four-dimensional (Euclidean) lattice. In this
formulation --- introduced by Wilson in 1974 \cite{Wilson:1974sk} ---
the theory becomes equivalent to a model in statistical mechanics and
can be studied numerically by Monte Carlo simulations \cite{binder}.
After over two decades \cite{Creutz:2003qy} of developments
in the methodology for the numerical study of QCD and
with present-day computers in the teraflops range, lattice-QCD simulations
are now able to provide quantitative predictions with errors of a
few percent. This means that these simulations will soon
become the main source of theoretical results for comparison
with experiments in high-energy physics \cite{Davies:2003ik}, enabling a much
more complete understanding of the physics of the strong force.

The study of lattice QCD constitutes a {\em Grand Challenge} computational
problem \cite{culler}. Consequently, lattice-QCD physicists
are natural users of high-performance computing
and have contributed to the development of supercomputer
technology itself. In fact, several research groups have built
QCD-dedicated computers, using parallel architecture.
Examples are the {\tt Hitachi/CP-PACS} machine at the
University of Tsukuba in Japan \cite{kanaya:2001aq},
the {\tt QCDSP} and {\tt QCDOC} machines at
Columbia University in the USA \cite{Mawhinney:2000fx,Boyle:2003mj},
and the {\tt APE} machines \cite{Bartoloni:1998gi,Ammendola:2002xt} at
various research centers in Italy and Germany. These computers
range from about 1 to 10 teraflops. In addition to these
large projects, many groups base their simulations on
clusters of workstations or personal computers (PC's)
\cite{Luscher:2001tx}, since
costs are much lower and maintenance is simpler.

In Brazil, the first PC cluster dedicated to lattice-QCD studies was
installed in 2001 at the Physics Department of the University of S\~ao
Paulo in S\~ao Carlos (IFSC--USP), as part of a FAPESP project.
The group is currently investigating the behavior of gluon and
ghost propagators in Landau gauge
\cite{Cucchieri:2001za,Bloch:2002we}, with the goal of
verifying Gribov's proposed mechanism for quark confinement
\cite{Gribov:1978wm,Zwanziger:1994dh}.
This study requires careful consideration of the infrared behavior
of these propagators, i.e.\ their behavior at small
momentum $p$ (typically $p \ll 1$ GeV).
Since the smallest non-zero momentum that
can be considered on a lattice is given by $p_{min} \approx 2 \pi /L$
--- where $L$ is the size of the lattice in physical units ---
it is clear
that one needs to simulate at very large lattice sizes in order to probe the
small-momentum limit. For example, to have $p_{min} \approx 0.06$ GeV
with a lattice spacing of about $0.17$ fm (i.e.\ physically relevant
values for small momentum and fine enough lattice\footnote{%
We note that 1 fm = $10^{-13}$ cm is approximately the size of a proton.
}),
one needs to simulate on a lattice with $140$ sites in each direction.
This is considerably more than what can be currently done in
QCD simulations. On the other hand, Gribov's predictions are also
valid for simpler cases of lattice gauge theories \cite{Zwanziger:1994dh},
such as 
three-dimensional two-color QCD with infinitely massive quarks,
i.e.\ (pure) $SU(2)$ lattice gauge theory in three dimensions.
This corresponds to considering the $SU(2)$
[instead of the $SU(3)$] gauge group
\cite{Wuki}, taking three (instead of the usual four)
space-time dimensions and
making the so-called quenched approximation \cite{rothe}.
In this case it was possible to simulate on $140^3$ lattices 
and to see clear evidence of Gribov's predicted behavior for
the gluon propagator \cite{Cucchieri:2003di}. This represents the largest
number of points per direction ever considered in lattice-gauge-theory
simulations. The study is currently being extended to even
larger lattices (up to $260^3$) aiming at a more quantitative
understanding of Gribov's confinement scenario.
The consideration of very large lattice sizes
requires parallelization and high efficiency of the code
in order to obtain good statistics in the Monte Carlo
simulation. Thus, an optimized parallel code is of great
importance.

\vskip 2mm
The purpose of the present paper is to describe
the implementation of the code
used in the study above, including a discussion
of speedup (at fixed and variable volume)
and efficiency.
The algorithms for these simulations and their
parallelization are briefly reviewed in
Section \ref{sec:algo}.
Our PC cluster is described in Section
\ref{sec:perf}, together with
the performance of the code.
Finally, in Section \ref{sec:concl} we comment on the
results and report our conclusions.

\section{The algorithms}
\label{sec:algo}

Lattice field theories are defined\footnote{%
One usually considers
units such that $\hbar = c = 1$, where $\hbar$ is the
Plank constant and $c$ is the speed of light in vacuum.
}
by a functional of the fields --- the action $S[U]$ --- which
determines the (unrenormalized) statistical weight
$e^{- S[U]}$ of a given field configuration
$\left\{ U \right\}$. All quantities of interest, called
{\em observables}, are computed as
weighted averages over the configurations, with the
weight function above. For a generic observable ${\cal O}[U]$
this average is defined as
\be
\langle {\cal O} \rangle \equiv \frac{\sum_{U}\,
 {\cal O}[U] \, e^{- S[U]}}{\sum_{U}\, e^{- S[U]}}
\:\mbox{.}
\label{eq:stataver}
\ee
In our case $\left\{ U \right\}$ is given by the
gluon field $U_{\mu}(\bx)$, which is a matrix defined
on each site $\bx$ and for each direction $\mu$ of the lattice.
The observables we consider are the gluon and ghost propagators.

The standard Wilson action for
$SU(2)$ lattice gauge theory in $d$ dimensions
is \cite{Wilson:1974sk}
\be
S\left[U\right] \equiv 
   \, \frac{\beta}{2} \sum_{\mu, \nu = 1}^{d}\,
      \sum_{\bx}\, \biggl\{ \, 1 \, - \frac{1}{2} \Tr P_{\mu \nu}
                    \, \biggr\}
\label{eq:action}
\:\mbox{,}
\ee
where the plaquette $P_{\mu \nu}$ is given by the product of the
gluon fields $U_{\mu}(\bx)$ around a closed $1 \times 1$ loop:
\be
P_{\mu \nu} \equiv
U_{\mu}(\bx) \; U_{\nu}(\bx + \bun_{\mu}) \;
U_{\mu}^{-1}(\bx + \bun_{\nu}) \; U_{\nu}^{-1}(\bx) 
\:\mbox{.}
\ee
Here, $U_{\mu}(\bx)$ are $SU(2)$ matrices, $\bx$ (with coordinates
$x_{\mu} = 1\mbox{,}\, 2\mbox{,}\,\ldots\mbox{,}\,N_{\mu}$) are sites on
a $d$-dimensional lattice with periodic boundary conditions
and $\bun_{\mu}$ is a unit vector in the positive $\mu$ direction.
The parameter $\beta$ controls the proximity to the
continuum limit.
The action in eq.\ \reff{eq:action} is invariant under the
so-called {\em local gauge transformation}
\be
U_{\mu}(\bx) \,\to\, U_{\mu}^{(g)}(\bx) \equiv
  g(\bx) \; U_{\mu}(\bx) \; g^{-1}(\bx + \bun_{\mu})
\:\mbox{,}
\label{eq:gtransf}
\ee
where $g(\bx)$ are general $SU(2)$ matrices.
Indeed, gauge theories
are systems with redundant dynamical variables, which
do not represent true dynamical degrees of freedom.
This implies that the objects of interest are not the gluon
fields $U_{\mu}(\bx)$
themselves, but rather the classes (orbits) of gauge-related fields
$U_{\mu}^{(g)}(\bx)$.
The elimination of such redundant gauge degrees of freedom is
often essential
for understanding and extracting physical information from these
theories. This is usually done by a method called 
{\bf gauge fixing}, in which a unique representative is chosen
on each gauge orbit \cite{Giusti:2001xf}.

For an $SU(2)$ matrix we shall use the parametrization
\be
g \, \equiv \, g_{0} \, \1 + i \bsig \cdot \bvg \, = \,
    \left( \begin{array}{rr}
          g_{0} + i g_{3} & g_{2} + i g_{1} \\
        - g_{2} + i g_{1} & g_{0} - i g_{3}
    \end{array} \right)
\:\;\mbox{,}
\label{eq:su2para}
\ee
where the components of $\bsig \equiv ( \sigma_{1} \mbox{,} \,
\sigma_{2} \mbox{,} \, \sigma_{3} ) $ are the three Pauli matrices
\cite{Wuki} and $ \cdot $ stands for scalar product.
Then, the adjoint of a matrix $g \in SU(2)$ is given by
\be
g^{\dagger} = g^{-1} = g_{0} \, \1 - i
\bsig \cdot \bvg
\:\mbox{.}
\ee
Also, note that the unitarity condition $\det g = 1$
(where $\det$ indicates the determinant of a matrix)
implies $g_{0}^2 + g_{1}^2 + g_{2}^2 + g_{3}^2 = 1$, namely an
$SU(2)$ matrix can be considered as a four-dimensional unit vector.
For the gluon field $U_{\mu}(\bx) \in SU(2)$
one usually writes
\be
U_{\mu}(\bx) \,=\, A_{0, \mu}(\bx) \, \1 + i \bsig \cdot \bum_{\mu}(\bx)
\:\mbox{.}
\label{eq:defA}
\ee

Our goal is to evaluate numerically
the gluon and ghost propagators
$D(k)$ and $G(k)$, defined in Section \ref{sec:prop}
[see eqs.\ \reff{eq:Dkdef} and \reff{eq:Gk}].
To this end one needs to (i)
produce a thermalized configuration $\{ U_{\mu}(\bx) \}$
by Monte Carlo simulation, (ii)
gauge fix this configuration, (iii) evaluate the propagators
using the gauge-fixed configuration. These steps are
described in detail in Sections \ref{subsec:ther},
\ref{sec:gfix}, \ref{sec:prop} and are 
schematically represented in the code below:

{\bf
\begin{verbatim}
main()
{
/* set parameters: beta, number of
   configurations NC, number of
   thermalization sweeps NT, etc. */
     read_parameters();
/* {U} is the link configuration */
     set_initial_configuration(U);

     for (int c=0; c < NC; c++) {
          thermalize(U,NT);
          gauge_fix(U,g);
          evaluate_propagators(U,D,G);
     }
}
\end{verbatim}
}

Note that the gauge-fixing step consists in finding
a gauge transformation $\{ g (\bx) \}$ 
[see eq.\ \reff{eq:gtransf}] leading to a given
gauge condition. In our case, since we are interested
in Gribov's predictions, we employ the so-called
{\bf Landau gauge}.

The general setup of our simulations is described in
Section \ref{sec:perf}.

\subsection{Thermalization}
\label{subsec:ther}

In (dynamic) Monte Carlo simulations \cite{binder}
the weighted configuration-space
average defined in \reff{eq:stataver} is substituted by a time average 
over successive realizations (i.e.\ configurations)
of the considered system, which evolves according
to a Markov process in the so-called Monte Carlo time.
Usually, the system is updated by sweeping over
all sites of the lattice and generating a new value for the field
at each site based on the conditional probability distribution
obtained by keeping all other field variables fixed. In the
{\em heat-bath} update, for example, this single-site distribution is
sampled exactly. In QCD simulations one considers only
effectively independent field configurations. This means that
one follows the system's evolution for a large enough number of
time steps such that a
statistically independent new configuration
is generated, discarding the
intermediate steps. Performing the Monte Carlo iterations
to obtain such independent field configurations is called
{\em thermalization}.

To thermalize the fields $\{ U_{\mu}(\bx) \}$ we use a
standard heat-bath algorithm \cite{Creutz:1980zw} accelerated
by {\em hybrid overrelaxation} \cite{Adler:1989gc}. This corresponds
to doing $n$ {\em micro-canonical} (or energy-conserving) update sweeps
over the lattice, followed by one local ergodic update
(a heat-bath sweep). As explained below,
the micro-canonical sweeps are important for a more
efficient sampling of the configuration space.
For the heat-bath update, one considers
the contribution of a single link variable
$U_{\mu}(\bx)$ to the Wilson action \reff{eq:action}.
This single-link action is given by
\be
S_{SL} \,=\, -\, \frac{\beta}{2} \, \Tr \,\left[\,
              U_{\mu}(\bx) \, H_{\mu}(\bx) \,\right]
   \,+\,\mbox{constant}
\:\mbox{,}
\label{eq:Ssl}
\ee
where the ``effective magnetic field'' $H_{\mu}(\bx)$ is defined as
\ba
H_{\mu}(\bx) \equiv
      \sum_{\nu \neq \mu} \!\!\!\!\!\!\! & & \!\!\!\!\! \left[ \,
              U_{\nu}(\bx + \bun_{\mu}) \,
U_{\mu}^{-1}(\bx + \bun_{\nu}) \, U_{\nu}^{-1}(\bx) \right. \nonumber \\
& & \!\!\!\!\!\!\!\!\!\!\!\!\!\!\!\!\!\!\!\!\!\!
    \!\!\!\!\!\!\!\!\!\!\!\!\!\!\!\! \left.+\, 
   U_{\nu}^{-1}(\bx - \bun_{\nu} + \bun_{\mu}) \,
U_{\mu}^{-1}(\bx - \bun_{\nu}) \, U_{\nu}(\bx  - \bun_{\nu})
  \,\right]
\:\mbox{.}
\label{eq:Hdef}
\ea
Since the matrix $H_{\mu}(\bx)$ is proportional
to an $SU(2)$ matrix, we can write it as
\be
H_{\mu}(\bx) \equiv {\cal N}_{\mu}(\bx) \; {\widetilde H}_{\mu}(\bx)
\label{eq:defhtilde}
\:\mbox{,}
\ee
with ${\widetilde H}_{\mu}(\bx) \in SU(2)$ and ${\cal N}_{\mu}(\bx)
\equiv \sqrt{\det H_{\mu}(\bx)}$. Then,
by using the invariance of the group measure under
group multiplication, one obtains the
heat-bath update \cite{Creutz:1980zw}
\be
U_{\mu}(\bx) \to V\,{\widetilde H}^{-1}_{\mu}(\bx)
\:\mbox{,}
\ee
where
the $SU(2)$ matrix $V \,=\, v_{0} \1 + i \bsig \cdot \bvv$ must
be generated by choosing $v_{0}$ according to the distribution
\be
\sqrt{\,1\,-\,v_{0}^{2}\,}\,\exp{\left[\,\beta\,{\cal N}_{\mu}(\bx)\,v_{0}
\,\right]}
   \,dv_{0}
\label{eq:u0distri}
\ee
and the vector $\bvv$ (which is normalized to
$\sqrt{ 1\,-\,v_{0}^{2} }\,$) pointing along a uniformly chosen
random direction in three-dimensional space.

The vector $\bvv$ can be easily generated. For example,
if we use cylindrical coordinates we may take
\ba
v_{1} &=& \sqrt{ \left(1 - \rho^{2} \right)
                 \left(1 - v_{0}^{2} \right)}
                 \,\cos{\phi} \\
v_{2} &=& \sqrt{ \left(1 - \rho^{2}\right)
                  \left(1 - v_{0}^{2} \right)}
                   \,\sin{\phi} \\
v_{3} &=& \sqrt{  1 - v_{0}^{2} }\, \rho
\ea
with $\rho$ uniformly distributed in $[ -1\mbox{,}\, 1 ]$ and
$\phi$ uniformly distributed in $[ 0\mbox{,}\,2\,\pi ]$.
On the contrary, the problem of generating $v_{0}$
according to the distribution \reff{eq:u0distri} is considerably
more involved. Three different rejection methods for this
purpose are considered in \cite[Appendix A]{Edwards:1992eg}.
Here, we use their algorithms
called method 1 and method 2, with a cutoff value of $2.0$
for the quantity $\beta\,{\cal N}_{\mu}(\bx)$,
namely method 1 is used when $\beta\,{\cal N}_{\mu}(\bx) < 2.0$
and method 2 is used otherwise.

In order to implement the heat-bath method
one also needs a (parallelized) random number generator.
Here we adopt the {\tt RANLUX} generator,
which is based on chaos theory \cite{Luscher:1994dy}.
More precisely, we use a double-precision
implementation of {\tt RANLUX} (version 2.1) with luxury
level set to 2.

Let us now consider the (deterministic) micro-canonical update,
used in the hybrid overrelaxed algorithm:
\be
U_{\mu}(\bx) \,\to\, {\widetilde H}^{- 1}_{\mu}(\bx)\,
\mbox{Tr}\,\left[\, U_{\mu}(\bx)\,{\widetilde H}_{\mu}(\bx)\,\right]
\,-\, U_{\mu}(\bx)
\label{eq:Udeterministico}
\:\mbox{.}
\ee
From formulae \reff{eq:Ssl} and \reff{eq:defhtilde} it is easy
to see that this update does not change the value of the
action $S_{SL}$. On the other hand, the step in
\reff{eq:Udeterministico} represents a large move in
configuration space \cite{Adler:1989gc}. Thus, one can alternate
micro-canonical sweeps of the lattice and heat-bath
updates in order to reduce the problem of critical slowing-down,
which afflicts Monte Carlo simulations of critical phenomena
\cite{Sokal:1989ea}.
The efficiency of the
hybrid overrelaxed algorithm may be optimized
by tuning the value of $n$,
i.e.\ the number of micro-canonical sweeps between two heat-bath sweeps.

\subsection{Landau gauge fixing}
\label{sec:gfix}

For a  given thermalized lattice configuration
$\{ U_{\mu}(\bx) \}$,
Landau gauge fixing is obtained by looking for a
gauge transformation $\{ g (\bx) \in SU(2) \}$ that brings the
functional
\be
{\cal E}_{U}[g] \equiv
- \sum_{\mu = 1}^{d} \sum_{\bx} \, \Tr \left[ \,
             g(\bx) \; U_{\mu}(\bx) \; g^{\dagger}(\bx + \bun_{\mu})
                        \, \right]
\label{eq:landau}
\ee
to a local minimum, starting from randomly chosen
$\{g(\bx)\}$ \cite{Giusti:2001xf}.
Thus, from the numerical point of view,
fixing the lattice Landau gauge is a minimization problem.
Here, we consider three different (iterative) gauge-fixing algorithms
\cite{Cucchieri:1996pn,Cucchieri:1997jm}:
the so-called {\em Cornell} (COR) method,
the {\em stochastic overrelaxation}
(SOR) algorithm and the {\em Fourier acceleration} (FA) algorithm.
Let us notice that the first two algorithms are based on local
updates for the matrices $ g(\bx)$ and have dynamic critical
exponent $z = 1$. This means that
the number of iterations required to achieve a given
accuracy in the minimization of the functional
${\cal E}_{U}[g]$ grows as a function of the
lattice side $N$ as $N^{d + 1}$, when
considering a symmetric lattice in $d$ dimensions. On the other hand,
the FA algorithm is based on a global update and has $z = 0$
[at least for sufficiently smooth field configurations
$\{ U_{\mu}(\bx) \}$].
Its computational work grows roughly as $N^{d}$.
Thus, even though the CPU time necessary to update a single-site
variable $g(\bx)$ is much smaller for the two local methods
than for the FA method, the latter should clearly be used when
considering very large values of $N$.

The update $g(\bx) \to g_{new}(\bx)$ for the COR and
the SOR methods
can be written in terms of local quantities --- i.e.\
quantities defined only in terms of
the site $\bx$ --- and of the matrix
\ba
h(\bx) \!\!\!& \equiv &\!\!\!
        \sum_{\mu = 1}^{d}
            \left[ \, U_{\mu}(\bx) \, g^{\dagger}(\bx + \bun_{\mu})
            \right. \nonumber \\
  & & \left. \qquad +\, U_{\mu}^{\dagger}(\bx - \bun_{\mu})
        g^{\dagger}(\bx - \bun_{\mu}) \right]
\:\mbox{.}
\label{eq:hdefi}
\ea

On the contrary, for the FA method one needs
to evaluate
\be
\bvu(\bx) \;=\;\left[\,\left(\,- \Delta\,\right)^{-1} \,
 \bvw\,\right]\!(\bx)
\:\mbox{,}
\label{eq:bvubis}
\ee
where $ {\mathrm w}(\bx) \!\equiv\!g(\bx) h(\bx)$, the three-dimensional
vector $\bvw$ is given by [see eq.\ \reff{eq:su2para}]
\be
2 \, i \,\bsig\cdot \bvw(\bx)  \equiv 
 {\mathrm w}(\bx) \,-\,{\mathrm w}^{\dagger}(\bx)
\ee
and $- \Delta$ is (minus) the lattice Laplacian, defined
for a general vector field $\bfv(\bx)$ as
\be
\!\left(-\Delta \bfv\right)\!(\bx) \equiv
\sum_{\mu = 1}^{d}\left[ 2 \bfv(\bx) - \bfv(\bx\! +\! \bun_{\mu})
           - \bfv(\bx\! -\! \bun_{\mu}) \right] 
\label{eq:laplaciano}
\:\mbox{.}
\ee
From eq.\ \reff{eq:bvubis} it is evident that the
FA method is actually a Laplacian preconditioning
algorithm.

Traditionally the inversion of the lattice Laplacian
is done using a fast Fourier transform (FFT), after
writing
\be
\left(\,- \Delta\,\right)^{-1} \,=\,
{\widehat F}^{-1}\, \frac{1}{p^2} \,{\widehat F}
\:\mbox{,}
\ee
where ${\widehat F}$ indicates the Fourier transform,
${\widehat F}^{-1}$ is its inverse and $p^{2}$
is the squared magnitude of the lattice momentum.
Alternatively \cite{Cucchieri:1998ew}, the inversion
of the Laplacian may be done using a multigrid (MG) algorithm or a
conjugate gradient (CG) method, avoiding the use of the FFT,
which has high communication costs in a parallel
implementation. Indeed, one obtains
\cite{Cucchieri:1998ew,Cucchieri:2003fb} the same convergence as
for the original algorithm (based on FFT), when using an accuracy
of about $10^{-3}$ for the MG or CG (iterative) inversion.
At the same time, the computational
cost of the new implementations is smaller than that of
the FFT-FA method when considering large lattice volumes.
This is true even for a non-parallelized code.
Moreover, the MG-FA
and CG-FA algorithms are well suited for vector
and parallel machines and they make the FA method more flexible,
i.e.\ it works equally well with any lattice side.\footnote{%
We note that the FFT is slightly less efficient for
lattice sides $N$ that are not powers of 2
\protect\cite{fftw}.
}
Here we do the inversion of the Laplacian
using a CG method preconditioned with red/black ordering
\cite{dongarra}. As stopping criterion we consider
$r_t / r_0 \leq 10^{-3}$, where $r_t$
is the magnitude of the CG residual after t iterations.
We note that, with this stopping criterion, one
can do the inversion in single
precision, even though the rest of the code is written in
double precision. This corresponds to a speed-up of almost
a factor 2 in the inversion.

Let us stress that the three algorithms considered above
require the tuning of a free parameter
in order to attenuate critical slowing-down, or equivalently in order
to reduce the computational work.
Notice however that
the CPU time necessary to update a single-site
variable $g(\bx)$ is essentially independent of the value of the
tuning parameter.

\subsubsection{Convergence of the gauge fixing}
\label{sec:converg}

Several quantities have been introduced in order to
check the convergence of Landau-gauge-fixing algorithms
\cite{Cucchieri:1996pn}. We consider here
\be
(\nabla A)^2 \, \propto \,  \sum_{\bx} \,
\sum_{b\, = 1}^{3} \, \Big[
    \left( \nabla \cdot A_{b} \right) (\bx) \Big]^{2}
\label{e2}
\:\mbox{,}
\ee
which is commonly used in numerical simulations, and
\be
\Sigma_Q \, \equiv \, \frac{1}{d} \,
  \sum_{\mu = 1}^{d} \, \frac{1}{3\,N_{\mu}}
\sum_{b\, = 1}^{3} \, \sum_{x_{\mu} = 1}^{N_{\mu}} \, \frac{
    \left[ Q_{b\mbox{,}\mu}(x_{\mu}) - {\overline Q}_{b\mbox{,}\mu} \,
      \right]^{2} }{ \left[ {\overline Q}_{b\mbox{,}\mu} \right]^{2} }
\:\mbox{,}
\label{eq:e6}
\ee
which provides a very sensitive test of the goodness of the
gauge fixing. Let us recall that
\be
\left(\nabla\cdot A_{b} \right)(\bx) \equiv
  \sum_{\mu = 1}^{d} \, \left[ A_{b\mbox{,}\mu} (\bx) -
                  A_{b\mbox{,}\mu} (\bx - \bun_{\mu}) \right]
\ee
is the lattice divergence of $A_{b\mbox{,}\mu}(\bx)$ [see
eq.\ \reff{eq:defA}]. We also define
\be
{\overline Q}_{b\mbox{,}\mu} \,
\equiv \, \frac{1}{N_{\mu}} \, \sum_{\bx_{\mu} = 1}^{N_{\mu}} \,
            Q_{b\mbox{,}\mu}(\bx_{\mu})
\: \mbox{,}
\ee
where the quantities
\be
\;\;\;\;Q_{b\mbox{,}\mu}(\bx_{\mu}) \, \equiv \, \sum_{\nu \neq \mu} \,
     \sum_{\bx_{\nu}} \, A_{b\mbox{,}\mu}(\bx)  
  \qquad \;\; \mu = 1\mbox{,}\ldots\,\mbox{,} d
\label{eq:charges}
\ee
are constant (i.e.\ independent of $\bx_{\mu}$) if the
Landau-gauge-fixing condition is satisfied.
The two quantities $(\nabla A)^2$ and $\Sigma_Q$
are expected to converge to zero
exponentially (and with the same exponent) as a function of the
number of gauge-fixing sweeps, even though their sizes may
differ considerably.

\subsection{Evaluation of the propagators}
\label{sec:prop}

A propagator of a field is a two-point function,
i.e.\ a correlation function between values of the field
at two different points in space-time \cite{lebellac}.
In quantum mechanics, the propagator
determines the evolution of the wave function of a system and,
for a particle, it gives the probability amplitude of 
going (i.e.\ propagating) from a point in space-time to another
\cite{sakurai}.
More generally, Green's functions (i.e.\ $n$-point
functions) carry all the information about the physical
and mathematical structure of a quantum field theory.
From this point of view,
two-point functions (propagators) are a theory's most basic
quantities and the gluon propagator
may be thought of as the most basic quantity of QCD.
The ghost propagator appears in the theory as a consequence
of the gauge-fixing procedure described above.

The gluon propagator is conveniently defined in
momentum space as
\be
D(\bk) \, \propto \, \sum_{\mu\mbox{,}\,b}\,\langle\,
| \sum_{\bx}\,A_{b\mbox{,}\mu}(\bx)\,
\exp{( 2 \pi i \bk \cdot \bx )}\,|^{2} \, \rangle
\:\mbox{.}
\label{eq:Dkdef}
\ee
Here, $b$ goes from $1$ to $3$ [when considering
the $SU(2)$ gauge group], $\mu$ goes from $1$ to $d$
($d = 3$ for three-dimensional space-time),
$k$ has components $k_{\mu}$ taking values
$k_{\mu}\,N_{\mu} =
0\mbox{,}\,1\mbox{,}\,\ldots \mbox{,}\, N_{\mu} - 1$ and
the field $A_{b\mbox{,}\mu}(x)$ is defined in eq.\ \reff{eq:defA}.
Note that $ \cdot $ stands for scalar product and
$ | \ldots | $ indicates the norm of a complex number.

The numerical evaluation of the ghost
propagator (in momentum space) is considerably more involved.
In fact, one has to calculate
\be
\;\; G(\bk) \propto
\sum_{\bx\mbox{,}\by} e^{-\bk\cdot (\bx - \by)}\,
\sum_{b}\,\langle\,
     \left(\,{\cal M}^{- 1}\,\right)_{b\,b}(\bx\mbox{,}\,
       \by\mbox{;}\,U)\,\rangle
\label{eq:Gk}
\:\mbox{,}
\ee
where the matrix ${\cal M}_{a b}(\bx\mbox{,}\, \by\mbox{;}\,U)$ is
a sparse matrix that depends on the gluon field $U_{\mu}(\bx)$.
(For an explicit definition of this matrix see eq.\ (B.18) in
\cite{Zwanziger:1994dh}.) Note that,
since the color indices $a$ and $b$ go from
$1$ to $3$ [for the $SU(2)$ group]
and if there are $N^d $ lattice sites,
the size of this matrix is $3 N^d \times 3 N^d$.

The inversion of the matrix ${\cal M}_{a b}(\bx\mbox{,}\, \by\mbox{;}\,U)$
can be done using a CG method with red/black ordering, as in the
case of the lattice Laplacian considered above for
gauge fixing. This part of
our code has not been parallelized yet, but we expect
to obtain a speedup comparable to the one obtained for the
lattice-Laplacian case.

\subsection{Parallelization}
\label{sec:par}

As said above, we need a parallelized code in order to
simulate at very large lattice sizes. We have
started by considering the {\tt QCDMPI} 
package \cite{qcdmpi}, which is based on
the work of Hioki \cite{hioki96}. The advantages of this
package are its portability and the
efficient way of evaluating the effective magnetic field
$H_{\mu}(\bx)$ --- also called the {\em staple} --- defined in
eq.\ \reff{eq:Hdef}. In particular, the extra memory space
required for communication is considerably reduced with
respect to previous implementations.
The original {\tt QCDMPI} code is written for pure $SU(3)$
lattice gauge theory in $d$ dimensions ($d \geq 2$)
and performs only the (heat-bath) thermalization step of the
simulation.
We have adapted the original code to the $SU(2)$ case and
improved the generation of
$v_{0}$ according to the distribution \reff{eq:u0distri}.
More precisely (see Section \ref{sec:algo}.{\em A}),
we have added method 1 and a more efficient
version of method 2.
At the same time, we have introduced the micro-canonical step,
the various Landau-gauge-fixing algorithms discussed
in Section \ref{sec:algo}.{\em B} (i.e.\ the COR,
SOR and CG-FA methods),
the calculation of the quantities $(\nabla A)^2$ and $\Sigma_Q$
for checking the convergence of the gauge fixing,
and the evaluation of the gluon propagator.
(As mentioned above, we have not yet parallelized the
evaluation of the ghost propagator.)

For the parallelization, we divide the lattice equally among
the nodes, i.e.\ we place $v= V / M$ sites of the lattice in
each node, where $V$ is the lattice volume and we use
$M$ nodes. Each node gets a contiguous block
of lattice sites. We will refer to this block of sites
as the {\em local lattice} in a node.
Note that, in general, not all directions $\mu$ of the
lattice are divided between different nodes and that, in order
to use a red/black ordering, the number of sites $v$ in each node
must be even.

Let us stress that communication is required for the evaluation
of the staple $H_{\mu}(\bx)$, for the
calculation of $h(\bx)$ [see eq.\ \reff{eq:hdefi}] and (in
the CG-FA method) for the inversion of the lattice Laplacian
[see eq.\ \reff{eq:laplaciano}]. Also, the evaluation of
the quantities $(\nabla A)^2$ and $\Sigma_Q$ (see Section
\ref{sec:algo}.B.1) requires some level of communication
in a parallel code, while for the gluon
propagator [see eq.\ \reff{eq:Dkdef}] one has to perform only a
sum over the whole lattice.
Since the expressions to be parallelized
involve at most quantities at
lattice sites that are nearest neighbors,
communication is required only for sites on the boundary
of the local lattice in a node. Moreover, simulations are
usually done in three or four dimensions, leading to a
high granularity due to the surface/volume effect.

All communications in \cite{hioki96} are carried out
using just two subroutines.
The first (called {\tt setlink}) sends data from a
node to the previous one in a given direction.
The second subroutine (called {\tt slidematrix})
sends data from a node to the next one (in
a given direction).
Clearly, these two routines are all we need in
order to perform the communications required
in our code.

\section{Performance}
\label{sec:perf}

As mentioned in the Introduction, our simulations were done on a
PC cluster at the IFSC-USP. The system has 16 nodes and a
server, all with 866 MHz Pentium III CPU. The nodes have
256 MB RAM memory (working at 133 MHz) and the operating
system is {\tt Debian GNU/Linux} (version 3.0r0).
The machines are connected with a 100 Mbps full-duplex network
through a {\tt 3COM} switch. All user directories are located
on the server, which has two {\tt SCSI} disks, and are
mounted by the nodes (using {\tt NFS}). The server is not used
for the computations. 

Our code (as well as the {\tt QCDMPI} package) is written
in {\tt FORTRAN 77} making use of {\tt MPI} for communication.
The code may be run for a general lattice dimension $d \geq 2$.
As said before, we consider here $d = 3$.
We use {\tt MPICH} (version 1.2.1-16) and the compiler {\tt g77}
(version 0.5.24).
The compilation has been done with the following four options
-march=pentiumpro 
-fomit-frame-pointer
-mpreferred-stack-boundary=2 
-O3.

\begin{table}[ht]
\begin{center}
\caption{Average
CPU-time (in $\mu s$) to
update a link variable $U_{\mu}(\bx)$ using heat-bath ($t_{hb}$)
or micro-canonical update ($t_{mc}$). Errors are one standard
deviation.}
\label{table-fixed-vol}
\begin{tabular}{c|c|c|c}
$M$ & Node topol.\ & $t_{hb}$ & $t_{mc}$ \\
\hline\hline
 1 & $  1 \times 1 \times 1 $ & 10.341(4) &  6.0190(1) \\ 
\hline
 2 & $  2 \times 1 \times 1 $ &  5.6(2)   &  3.2099(4) \\ 
\hline
 4 & $  2 \times 2 \times 1 $ &  2.78(2)  &  1.6958(3) \\ 
 4 & $  4 \times 1 \times 1 $ &  2.898(6) &  1.817(3)  \\ 
\hline
 8 & $  2 \times 2 \times 2 $ &  1.435(7) &  0.8881(3) \\ 
 8 & $  4 \times 2 \times 1 $ &  1.48(2)  &  0.9125(6) \\ 
 8 & $  8 \times 1 \times 1 $ &  1.9(1)   &  1.236(4)  \\ 
\hline
16 & $  4 \times 2 \times 2 $ &  0.758(7) &  0.4732(3) \\ 
16 & $  4 \times 4 \times 1 $ &  0.75(1)  &  0.4614(4) \\ 
16 & $  8 \times 2 \times 1 $ &  0.849(8) &  0.5677(3) \\ 
16 & $ 16 \times 1 \times 1 $ &  1.25(1)  &  0.6735(5) \\ 
\hline
\end{tabular}
\end{center}
\end{table}

We now describe the setup of a complete simulation.
As said in Section \ref{sec:algo}.{\em A}, we
generate statistically independent field configurations
to be used for the evaluation of the Monte Carlo average
of an observable, which provides an estimate of the
average in eq.\ \reff{eq:stataver}. In order to reduce
the statistical error (i.e.\ the Monte Carlo error) of
this estimate, one typically needs to produce hundreds
of such configurations. For each of them
there are two computational steps involved: the Monte Carlo
generation of a new independent field configuration and the
evaluation of the desired observables. The first step
is the thermalization, which in our case is done using the
hybrid overrelaxed algorithm. This usually requires hundreds
of sweeps of the lattice, corresponding to several
hours to produce a new configuration. The second step is
often even more time-consuming. 
In our case the evaluation of the observables (i.e.\
the propagators) is done only after the new
configuration has been gauge-fixed, by an iterative
minimization procedure. One usually needs
thousands of iterations in order to reach a
prescribed accuracy (for example $\Sigma_Q \leq 10^{-12}$,
required for a good quality of the gauge fixing).
The actual computation of the gluon propagator
requires a negligible time. (This is not true for the
ghost propagator.)
Consequently, for typical values of lattice volumes,
a complete simulation may take several months.
For example, in order to produce the (preliminary) data
reported in \cite{Cucchieri:2003di}, the code was
running on our PC cluster for almost three months.
The production of the corresponding final results
is expected to take almost one year of runs.
As an illustration, for $V = 140^3$ and $\beta = 6.0$
the average CPU-times (per configuration)
using 4 nodes are: about 8 hours for
thermalization and about 21 hours for gauge fixing
(using COR or SOR methods). We note that
the total CPU-time per configuration is not
appreciably affected by changes in the parameter $\beta$.
[More precisely, for smaller (respectively larger)
values of $\beta$ one spends a little less (resp.\
more) time for thermalization and a little more
(resp.\ less) time for gauge fixing.]
We plan to use the CG-FA method
in future production runs, to reduce the percentage of
time spent in gauge fixing.

We report below on some runs performed for testing
the speedup and efficiency of our code.

\subsection{Speedup at fixed volume}

We did tests for the various algorithms, considering a fixed
lattice volume $V = 64^3$, using 1, 2,
4, 8 and 16 nodes. Let us note that this lattice size
is relatively small. In fact, it can be simulated on a
single node (without parallelization) using less than
$20 \%$ of the memory (about $24 \%$ if one employs the CG-FA
method for gauge fixing). Thus,
since communications are proportional
to the surface area of the local lattice in a node,
these tests correspond to a worse situation than
the one we considered for our production runs
in \cite{Cucchieri:2003di}.

In Tables \ref{table-fixed-vol} and \ref{table-fixed-vol2}
we report the average CPU-time (in micro-seconds) necessary to
update a link variable $U_{\mu}(\bx)$ using a heat-bath ($t_{hb}$)
or a micro-canonical update ($t_{mc}$), 
and the time to update a site variable $g(\bx)$ using
the gauge-fixing
methods COR ($t_{cor}$), SOR ($t_{stoc}$) or CG-FA ($t_{cg}$).
These CPU-times are given for different values of
the number of nodes $M$ and different (three-dimensional) node topology.

\begin{table}[ht]
\begin{center}
\caption{Average CPU-time (in $\mu s$) to
update a site variable $g(\bx)$ using gauge-fixing
methods COR ($t_{cor}$), SOR ($t_{stoc}$) or CG-FA ($t_{cg}$).
Errors are one standard deviation.}
\label{table-fixed-vol2}
\begin{tabular}{c|c|c|c|c}
$M$ & Node topol.\ & $t_{cor}$ & $t_{stoc}$ & $t_{cg}$ \\
\hline\hline
 1 & $  1 \times  1 \times  1 $ &  5.606(2)  &  6.272(2)  & 253(1)    \\ 
\hline
 2 & $  2 \times  1 \times  1 $ &  2.9659(7) &  3.295(1)  & 136.2(2)  \\ 
\hline
 4 & $  2 \times  2 \times  1 $ &  1.560(1)  &  1.7232(5) &  70.9(3)  \\ 
 4 & $  4 \times  1 \times  1 $ &  1.6063(5) &  1.789(1)  &  73.6(3)  \\ 
\hline
 8 & $  2 \times  2 \times  2 $ &  0.819(1)  &  0.9011(4) &  37.5(1)  \\ 
 8 & $  4 \times  2 \times  1 $ &  0.833(1)  &  0.9175(7) &  37.2(1)  \\ 
 8 & $  8 \times  1 \times  1 $ &  1.0228(5) &  1.1133(4) &  43.5(1)  \\ 
\hline
16 & $  4 \times  2 \times  2 $ &  0.4383(3) &  0.4771(2) &  18.99(5) \\ 
16 & $  4 \times  4 \times  1 $ &  0.4315(4) &  0.4706(1) &  18.43(5) \\ 
16 & $  8 \times  2 \times  1 $ &  0.4870(6) &  0.5296(2) &  19.22(6) \\ 
16 & $ 16 \times  1 \times  1 $ &  0.5924(4) &  0.644(3)  &  27.29(5) \\ 
\hline
\end{tabular}
\end{center}
\end{table}

\subsection{Speedup at variable volume}

We also did tests at variable volume, considering five different
node topologies: $1 \times 1 \times 1$, $1 \times 1 \times 2$,
$1 \times 2 \times 2$, $2 \times 2 \times 2$ and
$2 \times 2 \times 4$, corresponding respectively
to $M = 1, 2, 4, 8$ and 16 nodes. For each node topology we
have simulated using three different lattice volumes $V$.
Results of these tests are reported in
Tables \ref{table-varia-vol} and \ref{table-varia-vol2}
for different numbers of nodes $M$ and for the various lattice volumes.
As can be seen from the first two columns in these tables,
the lattice volumes have been chosen so that the local lattice
volume $v = V / M$ is always given by one of the following cases:
$4^3, 16^3$ and $64^3$.
Let us note that this arrangement
is closer to what is usually considered for production runs.
In fact, when carrying out parallel simulations at increasingly
large lattice volumes on a PC cluster, it is preferable
to fill up the memory in each node before
increasing the number of nodes. (This reduces the
percentage of time spent in communication.)

\begin{table}[ht]
\begin{center}
\caption{Average
CPU-time (in $\mu s$) to
update a link variable $U_{\mu}(\bx)$ using heat-bath ($t_{hb}$)
or micro-canonical update ($t_{mc}$). Errors are one standard
deviation.}
\label{table-varia-vol}
\begin{tabular}{c|c|c|c}
$M$ & $V$ & $t_{hb}$ & $t_{mc}$ \\
\hline\hline 
 1 & $  4^3            $ &  4.76(1)   &  1.028(2)  \\ 
 1 & $ 16^3            $ &  8.05(6)   &  4.1821(9) \\ 
 1 & $ 64^3            $ & 10.383(5)  &  6.0186(1) \\ 
\hline
 2 & $  4^2 \times  8  $ &  9.19(2)   &  7.08(1)   \\ 
 2 & $ 16^2 \times 32  $ &  4.9(1)    &  2.7561(9) \\ 
 2 & $ 64^2 \times 128 $ &  5.39(7)   &  3.1184(2) \\ 
\hline
 4 & $  4 \times  8^2  $ &  8.1(2)    &  7.4(3)    \\ 
 4 & $ 16 \times 32^2  $ &  2.708(9)  &  1.6940(9) \\ 
 4 & $ 64 \times 128^2 $ &  2.768(3)  &  1.6175(3) \\ 
\hline
 8 & $  8^3            $ &  5.94(8)   &  5.02(6)   \\ 
 8 & $ 32^3            $ &  1.54(2)   &  1.0090(2) \\ 
 8 & $ 128^3           $ &  1.55(8)   &  0.8447(3) \\ 
\hline
16 & $  8^2 \times 16  $ &  5.8(6)    &  4.4(2)    \\ 
16 & $ 32^2 \times 64  $ &  0.763(2)  &  0.5111(3) \\ 
16 & $ 128^2 \times 256 $ &  0.77(1)  &  0.436(1) \\ 
\hline
\end{tabular}
\end{center}
\end{table}

\begin{table}[ht]
\begin{center}
\caption{Average CPU-time (in $\mu s$) to  
update a site variable $g(\bx)$ using gauge-fixing
methods COR ($t_{cor}$), SOR ($t_{stoc}$) or CG-FA ($t_{cg}$).
Errors are one standard deviation.}
\label{table-varia-vol2}
\begin{tabular}{c|c|c|c|c}
$M$ & $V$ & $t_{cor}$ & $t_{stoc}$ & $t_{cg}$ \\
\hline\hline
 1 & $  4^3             $ &  1.21(7)   &  1.60(1)   & 4.26(3)  \\ 
 1 & $ 16^3             $ &  3.90(1)   &  4.224(6)  & 19.82(4  \\ 
 1 & $ 64^3             $ &  5.606(2)  &  6.272(2)  & 253(1)   \\ 
\hline
 2 & $  4^2 \times  8   $ &  6.41(8)   &  6.79(4)   & 117.2(6) \\ 
 2 & $ 16^2 \times 32   $ &  2.413(5)  &  2.679(4)  & 31.38(6) \\ 
 2 & $ 64^2 \times 128  $ &  2.885(1)  &  3.2329(4) & 122(3)   \\ 
\hline
 4 & $  4 \times  8^2   $ &  6.2(2)    &  6.10(6)   & 125.1(2) \\ 
 4 & $ 16 \times 32^2   $ &  1.497(3)  &  1.571(3)  & 26.47(5) \\ 
 4 & $ 64 \times 128^2  $ &  1.4935(2) &  1.6634(3) & 63.7(7)  \\ 
\hline
 8 & $  8^3             $ &  4.53(6)   &  4.64(5)   & 96.1(9)  \\ 
 8 & $ 32^3             $ &  0.862(2)  &  0.927(1)  & 18.83(8) \\ 
 8 & $ 128^3            $ &  0.7620(2) &  0.8528(1) & 31.5(2)  \\ 
\hline
16 & $  8^2 \times 16   $ &  3.56(7)   &  4.4(2)    & 89.6(7)  \\ 
16 & $ 32^2 \times 64   $ &  0.4496(7) &  0.480(1)  & 14.25(4) \\ 
16 & $ 128^2 \times 256 $ &  0.3942(3) &  0.4385(2) & 17.6(5)  \\ 
\hline
\end{tabular}
\end{center}
\end{table}

\section{Results and conclusions}
\label{sec:concl}

The CPU-times reported above indicate that the parallelization
is rather good for the five algorithms considered
(the heat-bath and micro-canonical methods for thermalization,
and the COR, SOR and CG-FA methods for gauge fixing).
Also, the values for the speedup $S = t_1 / t_M$ (and the efficiency $E =
S / M$) are very similar for the five cases.
We therefore take averages over the five methods and report them
in Table \ref{efficiency}. In the fixed-volume case we present
results for node topologies $2 \times 2\times 1$, $\,2 \times 2\times 2$ and
$4 \times 4\times 1$ respectively for the cases $M = 4, 8$ and 16.
(This corresponds to the best performance for a given number of nodes.)
In the variable-volume case we consider results obtained
using the largest lattice volume for each node topology.
We clearly see that one obtains a good parallelization
even in the fixed-volume case and using a relatively
small lattice volume $V$. Indeed, the efficiency decreases
slowly when doubling the number of nodes and it is still about
0.84 at $M = 16$. As expected, the performance is better
for the variable-volume case when considering large values of the
local lattice volume $v$ in a node. As said above, this test
is closer to the situation usually considered in production
runs, with local lattice sizes using more than $50 \%$ of the
memory of each node.

\begin{table}[t]
\begin{center}
\caption{Average speedup and
efficiency.}
\label{efficiency}
\begin{tabular}{c|c|c|c|c}
 & \multicolumn{2}{c|}{Fixed volume} & \multicolumn{2}{c}{Variable volume} \\
         \cline{2-3} \cline{4-5} 
 & $S$ & $E$ & $S$ & $E$ \\
\hline\hline
$1 \to  2$ &  1.87(2) & 0.937(8) & 1.96(2) & 0.981(8) \\
$1 \to  4$ &  3.61(1) & 0.904(3) & 3.79(1) & 0.948(3) \\
$1 \to  8$ &  6.91(2) & 0.863(2) & 7.31(8) & 0.91(1)  \\
$1 \to 16$ & 13.38(6) & 0.836(3) & 14.5(3) & 0.88(1) \\
\hline
\end{tabular}
\end{center}
\end{table}

\begin{figure}[ht]
\caption{Code speedup $S$ at variable volume as
a function of the number of nodes $M$.}
\label{efficiency-plot}
\epsfxsize = 0.5\textwidth
\vspace{-4.0cm}
\hspace{-0.6cm}
\epsffile{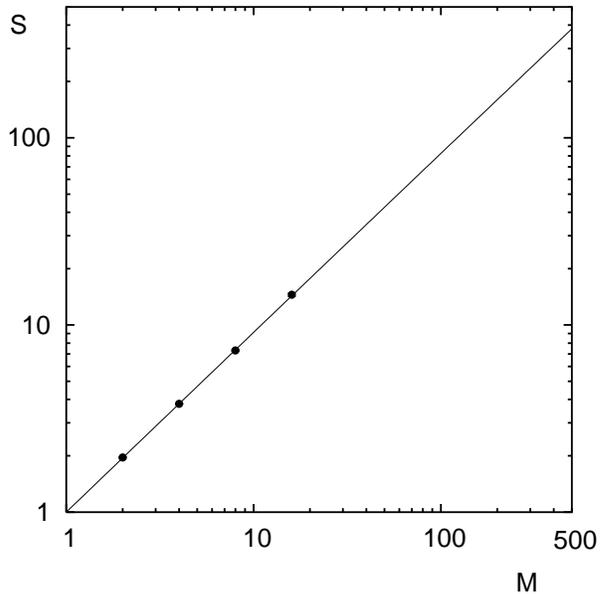}
\end{figure}

The data for the speedup at variable volume reported
in Table \ref{efficiency} are well fitted by the function
$S(M) = M \, (1 \,-\, c\,\log M)$ with $c = 0.038 \pm 0.001$. 
The corresponding plot is shown in Figure \ref{efficiency-plot}. 
This would mean a speedup of almost 400 for 512 nodes.

Note that the efficiency loss in going from 1 to $M$ nodes at
variable volume $V$ and small $v$ is due to the effect of
node communication over the usage of the cache memory.

We conclude that the parallelization of our code
works very well and that simulations of this type
are very viable on a PC cluster. Let us point out again
the main results of the parallel implementation described
in this paper. We have adapted and extended the package
{\tt QCDMPI}, which shows good parallelization but performs
only thermalization of the gluon fields. The resulting
code employs a more efficient thermalization algorithm
and also performs gauge fixing and evaluation of propagators,
with essentially the same parallel performance as the
original package.

For the future we plan to (i) parallelize the codes
for the evaluation of the ghost propagator and
for the MG-FA gauge-fixing method, (ii) improve the
usage of the cache memory and (iii) introduce overlap
of computation and communication.

\section*{Acknowledgements}
We thank Martin L\"uscher for sending
us the latest version of the 
random number generator {\tt RANLUX}.
The research of A.C.\ and T.M.\ is
supported by FAPESP (Project No. 00/05047-5).
A.T.\ thanks CNPq for financial support.

\bibliographystyle{ieee}
\bibliography{attilio}

\end{document}